\documentclass[prl,showpacs,twocolumn,aps,superscriptaddress,a4paper]{revtex4-1}

\usepackage{pst-node}
\usepackage{amsmath}
\usepackage{graphicx}
\usepackage{latexsym}
\usepackage{amsfonts}
\usepackage{amssymb}
\usepackage{graphicx}
\usepackage{rotating}
\usepackage{bbm}
\usepackage{cancel}
\usepackage[normalem]{ulem}

\DeclareGraphicsExtensions{.eps,.pdf,.gif,.jpg,.png}

\newcommand{\be}{\begin{equation}}
\newcommand{\ee}{\end{equation}}
\newcommand{\bea}{\begin{eqnarray}} 
\newcommand{\eea}{\end{eqnarray}}
\newcommand{\nn}{\nonumber}

\begin{document}  

\def\gC{\mbox{\boldmath $C$}}
\def\gZ{\mbox{\boldmath $Z$}}
\def\gR{\mbox{\boldmath $R$}}
\def\gN{\mbox{\boldmath $N$}}
\def\gG{\mbox{\boldmath $G$}}
\def\green{\mbox{\boldmath ${\cal G}$}}
\def\grn{\mbox{${\cal G}$}}
\def\gH{\mbox{\boldmath $H$}}
\def\bA{{\bf A}}
\def\bJ{{\bf J}}
\def\bG{{\bf G}}
\def\bF{{\bf F}}
\def\bH{\mbox{\boldmath $H$}}
\def\bQ{\mbox{\boldmath $Q$}}
\def\bgS{\mbox{\boldmath $\Sigma$}}
\def\bT{\mbox{\boldmath $T$}}
\def\bU{\mbox{\boldmath $U$}}
\def\bV{\mbox{\boldmath $V$}}
\def\bgG{\mbox{\boldmath $\Gamma$}}
\def\bgL{\mbox{\boldmath $\Lambda$}}
\def\ubG{\underline{{\bf G}}}
\def\ubH{\underline{{\bf H}}}
\def\ubQ{\underline{{\bf Q}}}
\def\ubS{\underline{{\bf S}}}
\def\ubg{\underline{{\bf g}}}
\def\ubq{\underline{{\bf q}}}
\def\ubp{\underline{{\bf p}}}
\def\ubgS{\underline{{\bf \Sigma}}}
\def\bge{\mbox{\boldmath $\epsilon$}}
\def\bgD{{\bf \Delta}}

\def\bDelta{\mbox{\boldmath $\Delta$}}
\def\bcalE{\mbox{\boldmath ${\cal E}$}}
\def\bcalF{\mbox{\boldmath ${\cal F}$}}
\def\bcalG{\mbox{\boldmath $G$}}
\def\ubcalG{\mbox{\underline{\boldmath $G$}}}
\def\callG{\mathcal{G}}
\def\callA{\mathcal{A}}
\def\callT{\mathcal{T}}
\def\ubcalA{\mbox{\underline{\boldmath $A$}}}
\def\ubcalB{\mbox{\underline{\boldmath $B$}}}
\def\ubcalC{\mbox{\underline{\boldmath $C$}}}
\def\ubcalg{\mbox{\underline{\boldmath $g$}}}
\def\ubcalH{\mbox{\underline{\boldmath $H$}}}
\def\bcalK{\mbox{\boldmath $K$}}
\def\ubcalK{\mbox{\underline{\boldmath $K$}}}
\def\bcalV{\mbox{\boldmath ${\cal V}$}}
\def\ubcalV{\mbox{\underline{\boldmath $V$}}}
\def\bcalU{\mbox{\boldmath ${\cal U}$}}
\def\ubcalz{\mbox{\underline{\boldmath $z$}}}
\def\bcalQ{\mbox{\boldmath ${\cal Q}$}}
\def\ubcalQ{\mbox{\underline{\boldmath $Q$}}}
\def\ubcalP{\mbox{\underline{\boldmath $P$}}}
\def\bSS{\mbox{\boldmath $S$}}
\def\ubcalS{\mbox{\underline{\boldmath $S$}}}
\def\bff{\mbox{\boldmath $f$}}
\def\bg{\mbox{\boldmath $g$}}
\def\bh{\mbox{\boldmath $h$}}
\def\bk{\mbox{\boldmath $k$}}
\def\bq{\mbox{\boldmath $q$}}
\def\bp{\mbox{\boldmath $p$}}
\def\br{\mbox{\boldmath $r$}}
\def\bt{\mbox{\boldmath $t$}}
\def\ubh{\mbox{\underline{\boldmath $h$}}}
\def\ubt{\mbox{\underline{\boldmath $t$}}}
\def\ubk{\mbox{\underline{\boldmath $k$}}}
\def\ua{\uparrow}
\def\da{\downarrow}
\def\a{\alpha}
\def\b{\beta}
\def\g{\gamma}
\def\G{\Gamma}
\def\d{\delta}
\def\D{\Delta}
\def\e{\epsilon}
\def\ve{\varepsilon}
\def\z{\zeta}
\def\h{\eta}
\def\th{\theta}
\def\vth{\vartheta}
\def\k{\kappa}
\def\l{\lambda}
\def\L{\Lambda}
\def\m{\mu}
\def\n{\nu}
\def\x{\xi}
\def\X{\Xi}
\def\p{\pi}
\def\P{\Pi}
\def\r{\rho}
\def\bgr{\mbox{\boldmath $\rho$}}
\def\s{\sigma}
\def\us{\mbox{\underline{\boldmath $\sigma$}}}
\def\ubgm{\mbox{\underline{\boldmath $\mu$}}}
\def\S{\Sigma}
\def\ubcgS{\mbox{\underline{\boldmath $\Sigma$}}}
\def\t{\tau}
\def\f{\phi}
\def\vf{\varphi}
\def\F{\Phi}
\def\c{\chi}
\def\k{\kappa}
\def\w{\omega}
\def\W{\Omega}
\def\Q{\Psi}
\def\q{\psi}
\def\de{\partial}
\def\inf{\infty}
\def\ra{\rightarrow}
\def\bra{\langle}
\def\ket{\rangle}
\def\bbra{\langle\langle}
\def\kket{\rangle\rangle}
\def\grad{\mbox{\boldmath $\nabla$}}
\def\no{\bf 1}
\def\ze{\bf 0}
\def\uno{\underline{\bf 1}}
\def\zero{\underline{\bf 0}}

\def\dr{{\rm d}}
\def\bj{{\bf j}}
\def\br{{\bf r}}
\def\bz{\bar{z}}
\def\bart{\bar{t}}

\title{Nonequilibrium Anderson model made simple with density functional 
theory}

\author{S. Kurth}
\affiliation{Nano-Bio Spectroscopy Group and European Theoretical Spectroscopy 
Facility (ETSF), Dpto. de F\'{i}sica de Materiales,
Universidad del Pa\'{i}s Vasco UPV/EHU, Av. Tolosa 72, 
E-20018 San Sebasti\'{a}n, Spain}
\affiliation{IKERBASQUE, Basque Foundation for Science, Maria Diaz de Haro 3, 
E-48013 Bilbao, Spain}

\author{G. Stefanucci}
\affiliation{Dipartimento di Fisica, Universit\`{a} di Roma Tor Vergata,
Via della Ricerca Scientifica 1, 00133 Rome, Italy; European Theoretical 
Spectroscopy Facility (ETSF)}
\affiliation{INFN, Sezionde di Roma Tor Vergata, Via della Ricerca 
Scientifica 1, 00133 Rome, Italy}

\begin{abstract}
The single-impurity Anderson model is studied within the i-DFT framework, a 
recently proposed extension of density functional theory (DFT) for the description 
of electron transport in the steady state. i-DFT is designed to give 
both the steady current and density at the impurity, and 
it requires the knowledge of the exchange-correlation (xc) bias and on-site 
potential (gate). In this work we construct
an approximation for both quantities 
which is accurate in a wide range of temperatures, gates and biases, thus 
providing a simple and unifying framework to calculate the differential 
conductance at {\em negligible} computational 
cost in different regimes. Our results mark 
a substantial advance for DFT and may inform the construction of functionals 
applicable to other correlated systems. 
\end{abstract}


\maketitle

The description of strongly correlated systems in and, particularly, out of 
equilibrium is a challenge for any theoretical method. Density functional 
theory (DFT), despite its many successes in the {\em ab-initio} description 
of atoms, molecules, and solids, is certainly not the first method which comes 
to mind to tackle strong electronic correlation. In recent years, however, 
it has been realized that effects of strong correlation may indeed be within 
reach of the DFT framework 
\cite{CapelleCampo:13,LimaSilvaOliveiraCapelle:03,MaletGoriGiorgi:12,MirtschinkSeidlGoriGiorgi:13,MosqueraWasserman:14,HodgsonRamsdenChapmanLillystoneGodby:13,HodgsonRamsdenGodby:16,LorenzanaYingBrosco:12,YingBroscoLorenzana:14,sf.2008}. 
For instance, the 
Kondo plateau in the zero-bias conductance may already be captured at the 
level of standard Landauer theory combined with DFT 
\cite{Lang:95,tgw.2001,bmots.2002} provided that an accurate 
exchange-correlation (xc) potential is used~\cite{sk.2011,blbs.2012,tse.2012}. 
Similarly, it has been shown how the description of Coulomb blockade 
can be achieved within a DFT framework both in the zero-bias 
limit~\cite{ks.2013,LiuBurke:15,YangPerfettoKurthStefanucciDAgosta} as well 
as at finite bias~\cite{StefanucciKurth:15}.

The single-impurity Anderson model (SIAM)~\cite{Anderson:61} is the minimal model 
for the description of transport through a correlated system. Naturally, it 
has been studied with a wealth of techniques, especially in recent 
years. An incomplete list includes the time-dependent density matrix 
renormalization group~\cite{HeidrichMeisnerFeiguinDagotto:09}, 
functional renormalization group (fRG) in the 
linear~\cite{KarraschMedenSchoenhammer:10} and non-linear 
regimes~\cite{EckelHeidrichJakobsThorwartPletyokhovEgger:10,
JakobsPletyukhovSchoeller:10}, the numerical renormalization group 
(NRG)~\cite{IzumidaSakaiSuzuki:01,IzumidaSakai:05}, diagrammatic many-body 
methods~\cite{ThygesenRubio:07,
UimonenKhosraviStanStefanucciKurthLeeuwenGross:11}, and Quantum Monte Carlo 
(QMC) techniques~\cite{WernerOkaMillis:09,WernerOkaEcksteinMillis:10}. 
A recent comparative study~\cite{EckelHeidrichJakobsThorwartPletyokhovEgger:10} 
shows the level of agreement reached between some of these methods, giving 
confidence that their results can be considered as accurate reference. 

In the present work we exploit state-of-the-art reference values of the 
nonequilibrium SIAM to construct an accurate DFT functional 
allowing for the calculation of 
density and current at {\em negligible} computational 
cost for arbitrary interaction strength. 
The DFT results are shown to reproduce previously published differential 
conductances in a wide range of temperatures, on-site potentials and biases  
with very high accuracy. Our functional does not only provide a fast 
solution of the nonequilibrium SIAM but also offers an alternative perspective 
on how to attack more complicated models and/or other physical situations 
such as, e.g., time-dependent correlated transport.

We work in the  i-DFT framework, a recently proposed 
extension of DFT, designed to study open systems in the 
steady state~\cite{StefanucciKurth:15}. i-DFT
establishes a one-to-one map between the steady density $n({\bf r})$ of the open 
system and the steady current $I$
on one hand 
and the external potential (gate) and bias  on the other hand. In the spirit 
of DFT there exists an open Kohn-Sham (KS) system of non-interacting 
electrons with the same $n({\bf r})$ and $I$ as the 
interacting system. In the KS system the Hartree-xc (Hxc) contribution to the 
gate  and the xc contribution to the bias are functionals 
of $n({\bf r})$ and $I$. 

{\em i-DFT for the SIAM --}
In the SIAM the open system consists of a single-impurity (hence $n$ 
is the same as the total number of particles $N$) with on-site repulsion $U$ 
between opposite spin electrons and with energy independent tunneling rate 
$1/\g$ between the impurity and the left/right ($L/R$) electrodes. Denoting 
by $V$ the external bias, i-DFT leads to two coupled 
self-consistent KS equations for the steady density $N$ and 
current $I$ (hereafter $\int\equiv \int \frac{d\w}{2\p}$):
\begin{subequations}
\be
N= \sum_{\a=L,R}\int f\left(\w+s_{\a}\frac{V+V_{\rm xc}}{2}\right) A_s(\w) \;,
\label{ksn}
\ee
\be
I= \frac{\g}{2} \sum_{\a=L,R}\int f\left(\w+s_{\a}\frac{V+V_{\rm
xc}}{2}\right)s_{\a}A_s(\w)
\label{ksi}
\ee
\label{ks}
\end{subequations}

\noindent
where $s_{R/L}=\pm$ and $f(\w)=1/(1 + e^{\beta (\w - \mu)})$ is the Fermi function 
at inverse temperature $\beta=1/T$ and chemical potential $\mu$. Furthermore,  
\be
A_s(\w) = \frac{\g}{(\w - v_s)^2 + \frac{\g^2}{4}}
\label{ks_spectral}
\ee
is the KS spectral function with KS gate  
$v_s=v + v_{\rm Hxc}$,  $v$ being the external gate. 
Both the Hxc gate $v_{\rm Hxc}=v_{\rm Hxc}[N,I]$ and the xc bias
$V_{\rm xc}=V_{\rm xc}[N,I]$ are functionals of the steady density and current, 
and need to be approximated in practice. 

Equations~(\ref{ks}) are the basic self-consistency conditions of 
the i-DFT approach. They can also be used to 
derive an expression for the finite-bias differential conductance. The right-hand sides of 
Eqs.~(\ref{ks}) depend on $V$ both explicitly through the Fermi 
functions  and implicitly through 
$N$ and $I$ (which enter as arguments of the xc 
potentials). 
Differentiation of Eqs.~(\ref{ks}) 
with respect to $V$ leads to a linear system of  
coupled equations for $\frac{{\rm d} N}{{\rm d}V}$ and 
$\frac{{\rm d} I}{{\rm d}V}$ which can easily be solved. Since 
we are mainly concerned with differential conductances, we only give the 
explicit solution for this quantity 
\be
\frac{{\rm d} I}{{\rm d}V} = \frac{1}{D} \left( \frac{G_- + G_+}{2} + 
\frac{4}{\g} G_- G_+ \frac{\partial v_{\rm Hxc}}{\partial N} \right)
\label{dIdV}
\ee
where the denominator is defined as 
\bea
D &\equiv& 1 - \frac{1}{\g}(G_- - G_+) \frac{\partial V_{\rm xc}}{\partial N} 
+ \frac{2}{\g} (G_- + G_+) \frac{\partial v_{\rm Hxc}}{\partial N} \nn \\
&& - \frac{1}{2} (G_- + G_+) \frac{\partial V_{\rm xc}}{\partial I} 
+ (G_- - G_+) \frac{\partial v_{\rm Hxc}}{\partial I} \nn \\
&& + \frac{4}{\g} G_- G_+ \left( \frac{\partial v_{\rm Hxc}}{\partial I} 
\frac{\partial V_{\rm xc}}{\partial N} - \frac{\partial v_{\rm Hxc}}{\partial N} 
\frac{\partial V_{\rm xc}}{\partial I} \right)
\eea
and 
\be
G_{\pm} \equiv -\frac{\g}{2} \int f'\left(\w \pm \frac{V+V_{\rm xc}}{2}\right) 
A_s(\w) \;.
\label{G+-}
\ee
For given external gate $v$ and bias $V$, Eq.~(\ref{dIdV}) has to be evaluated 
at the self-consistent values of $N$ and $I$ found by solving 
Eqs.~(\ref{ksn}) and (\ref{ksi}). Interestingly, at zero bias and 
arbitrary gate $v$ we have 
$\frac{\partial v_{\rm Hxc}}{\partial I}\vert_{I=0}=
\frac{\partial V_{\rm xc}}{\partial N}\vert_{I=0}=0$
and $G_{-}=G_{+}=G_{s}(v)$, where $G_{s}(v)$ is the zero-bias 
KS conductance. 
It is straightforward to show that in this case Eq.~(\ref{dIdV}) 
simplifies to \cite{StefanucciKurth:15}
\be
G(v)\equiv \left.\frac{{\rm d} I}{{\rm d}V}\right|_{V=0}
=\frac{G_{s}(v)}{1-G_{s}(v)\left.\frac{\de V_{\rm xc}}{\de
I}\right|_{I=0}} \;.
\label{idftG0}
\ee
Similarly, at the particle-hole (ph) symmetric point $v=-U/2$ (hence $N=1$) 
and arbitrary bias $V$, we have 
$\frac{\partial v_{\rm Hxc}}{\partial I}\vert_{N=1}=
\frac{\partial V_{\rm xc}}{\partial N}\vert_{N=1}=0$. 
Furthermore, since $v_{\rm Hxc}[1,I]=U/2$ the KS spectral function 
is even in $\w$ and therefore $G_{-}=G_{+}=G_{{\rm ph},s}(V)$, where 
$G_{{\rm ph},s}(V)$ is the 
finite bias  KS conductance at the ph symmetric point. 
Then Eq.~(\ref{dIdV}) reduces to
\be
G_{\rm ph}(V)\equiv \left.\frac{{\rm d} I}{{\rm d}V}\right|_{v=-U/2}
=\frac{G_{{\rm ph},s}(V)}{1-G_{{\rm ph},s}(V)\left.\frac{\de V_{\rm xc}}{\de
I}\right|_{N=1}} \;.
\label{idftGph}
\ee

\begin{figure}[tbp]
\includegraphics[width=0.47\textwidth]{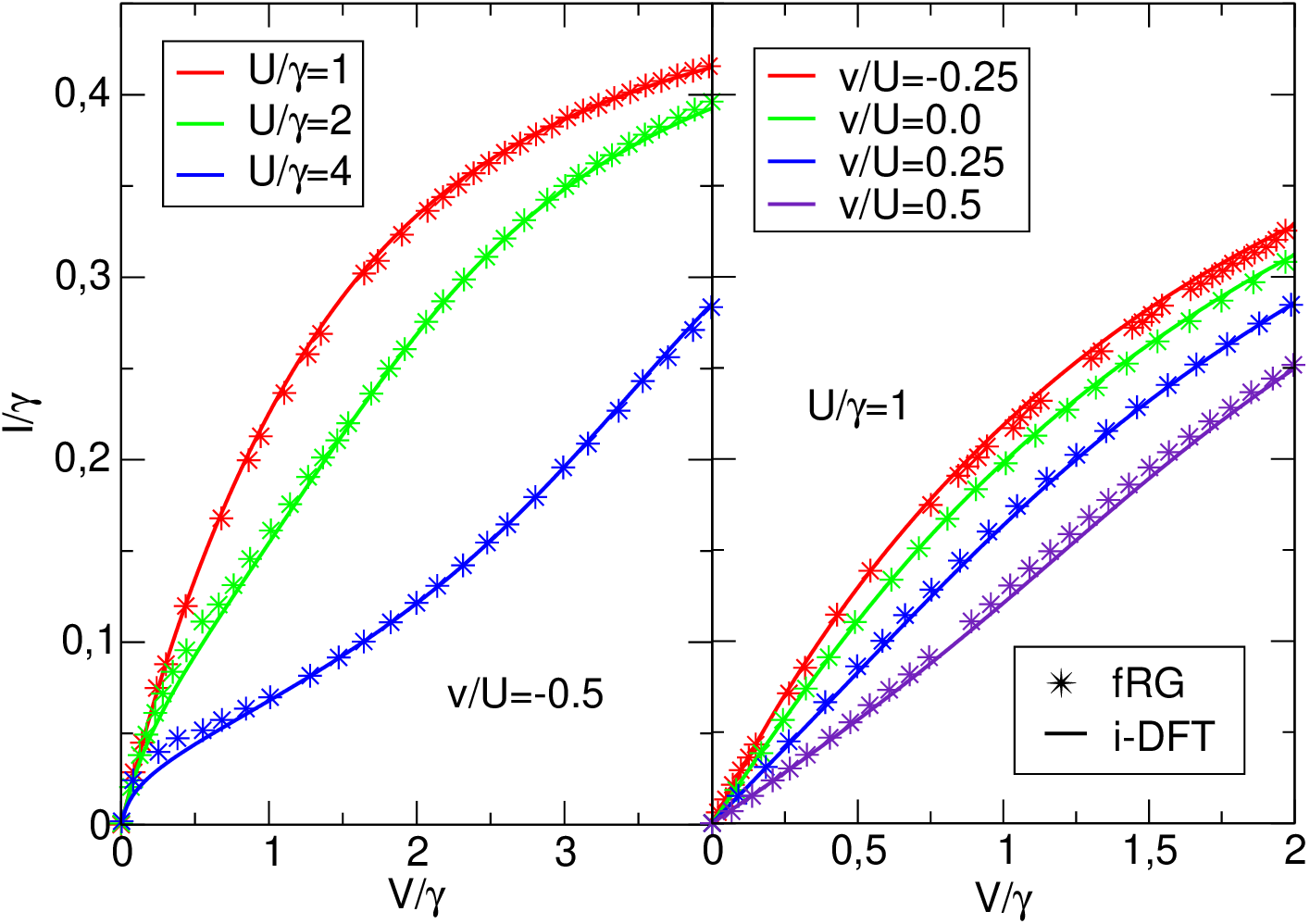}
\caption{Comparison between fRG and i-DFT $I$-$V$ characterisitcs 
at zero temperature. Left: at the ph 
symmetric point $v=-U/2$ for different $U/\g$. Right: at 
fixed $U/\g=1$ for different $v/U$. fRG results 
from Ref.~\protect\citenum{EckelHeidrichJakobsThorwartPletyokhovEgger:10}.}
\label{compare_IV_frg_fig}
\end{figure}

{\em xc potentials at zero temperature --}
In order to use the i-DFT formulas 
we need an approximation for $v_{\rm Hxc}$ and $V_{\rm xc}$. 
In Ref.~\citenum{StefanucciKurth:15} we 
showed that the Coulomb blockade diamond is correctly described by 
the (H)xc potentials 
\begin{subequations}
\bea
\tilde{v}_{\rm Hxc}[N,I]&=&\frac{U}{4}\sum_{s=\pm}\!\left[1+\frac{2}{\p}\,
{\rm atan}\left(\frac{N+\frac{s}{\g}I-1}{\l_{1} W_{0}}\right)\right]
\label{xcgate1}
\\
\tilde{V}_{\rm xc}[N,I]&=&-U\sum_{s=\pm}\frac{s}{\p}\,
{\rm atan}\left(\frac{N+\frac{s}{\g}I-1}{\l_{1} W_{0}}\right)\quad\quad
\label{xcbias1}
\eea
\label{xcpots1}
\end{subequations}

\noindent
where $W_{0}=0.16\g/U$ and the fitting parameter $\l_{1}$ was chosen to 
be $\l_{1}=1$. 
The essential property of the (H)xc
potentials of Eq. (8) are step-like features occuring at
the lines $N\pm I-1=0$ in the $N$-$I$ plane. 
Unfortunately, these potentials miss the Kondo plateau in $G(v)$ found at 
zero temperature. 
In fact, at $T=0$ we have  
$G=G_{s}$~\cite{mera-1,mera-2} and the Kondo 
plateau stems from the KS conductance 
alone (provided that the exact Hxc 
gate is used)~\cite{sk.2011,blbs.2012,tse.2012}. 
Although Eq.~(\ref{xcgate1}) at $I=0$ well approximates 
the exact $v_{\rm Hxc}$ (exhibiting a smeared step of height $U$ at 
half filling), we see from Eq.~(\ref{idftG0}) that 
$\frac{\de V_{\rm xc}}{\de I}\big\vert_{I=0}$ 
needs to vanish for the equality $G=G_{s}$ 
to hold. This is not the case for the approximation in Eq.~(\ref{xcbias1}). 
To incorporate the Kondo physics in the functionals we have to make sure 
that (a) the correction to $G_{s}$ vanishes and (b) the Hxc gate 
at zero current is as accurate as possible. Both requirements
can be  satisfied with the following ansatz:
\begin{subequations}
\bea
v_{\rm Hxc}[N,I]&=&\!\left( 1- \tilde{a}^{(0)}[I]\right) \!\tilde{v}_{\rm Hxc}[N,I]
+ \tilde{a}^{(0)}[I] v_{\rm Hxc}^{(0)}[N] \quad\;\;\;
\label{xcgate_zero_t}
\\
V_{\rm xc}[N,I]&=& \!\left( 1- \tilde{a}^{(0)}[I]\right)\! \tilde{V}_{\rm xc}[N,I]
\label{xcbias_zero_t}
\eea
\label{xcpots_zero_t}
\end{subequations}

\noindent
where $v_{\rm Hxc}^{(0)}[N] $ is the parametrization of the $T=0$
Hxc gate of Ref.~\citenum{blbs.2012}. There are a few 
constraints which restrict the choice of the function 
$\tilde{a}^{(0)}$. By 
symmetry, $\tilde{a}^{(0)}$ should be an even function of the current and, 
for $\frac{\de V_{\rm xc}}{\de 
I}\big\vert_{I=0}$ to vanish,
its value at vanishing current should be unity. 
Furthermore, the effect of $\tilde{a}^{(0)}$ should fade out as the current 
increases since the (H)xc potentials of Eq.~(\ref{xcpots1}) already 
give the 
physically correct picture at finite current. 
Here we choose the following form satisfying all these conditions
\be
\tilde{a}^{(0)}[I]= 1 - 
\left[ \frac{2}{\pi} \arctan\left( \frac{I}{\g W_0} \right) 
\right]^2 \;.
\label{gfunc_zero_t}
\ee 
Equations~(\ref{xcpots_zero_t}) and~(\ref{gfunc_zero_t}) 
completely specify the zero-temperature (H)xc potentials 
once a value of $\l_{1}$ in Eqs.~(\ref{xcpots1}) is 
chosen. In the left panel of Fig.~\ref{compare_IV_frg_fig} 
we show that for $\l_{1}=2$ the i-DFT $I$-$V$ characteristics
at the ph symmetric point is on top of the 
fRG results~\cite{EckelHeidrichJakobsThorwartPletyokhovEgger:10} 
in a wide bias window {\em and} for 
various values of $U/\g$. 
The value $\l_{1}=2$ performs well 
even away from  the ph symmetric point, thus supporting the general validity 
of the functional forms. 
In the right panel of Fig.~\ref{compare_IV_frg_fig} we compare the 
$I$-$V$ 
characteristics from fRG and i-DFT for various gates at a fixed 
value of $U/\g=1$; the agreement is excellent. We emphasize that in addition to 
the conceptual simplicity i-DFT is also numerically very efficient: the 
self-consistent solution of 
Eqs.~(\ref{ks}) is so fast that the calculation of one $I$-$V$ 
characteristics requires less than a CPU second. 

{\em xc-potentials at finite temperature --}
We now turn to the construction of the xc potentials at finite
temperatures. At the ph symmetric point the zero-bias 
conductance $G_{\rm ph}(0)$ is known to be a universal function 
$G_{\rm univ}(T/T_K)$ of the ratio between $T$ and the Kondo 
temperature $T_K$~\cite{AleinerBrouwerGlazman:02} which 
is given by~\cite{JakobsPletyukhovSchoeller:10}
\be
T_K = \frac{4}{\pi}\sqrt{U \g} 
\exp\left( 
-\frac{\pi}{4}\left(\frac{U}{\g}-\frac{\g}{U}\right)\right) .
\ee
The function $G_{\rm univ}$ has 
been calculated using the NRG method in 
Ref.~\citenum{costi:00}. To reproduce this 
universal behavior we keep the form  in Eqs.~(\ref{xcpots_zero_t}) 
except for replacing $W_{0}$ with a temperature dependent $W(T)$ and 
$\tilde{a}^{(0)}[I]$ with a temperature-dependent 
functional of $N$ and $I$: 
\be
a^{(T)}[N,I] = b^{(T)}[N] \tilde{a}^{(T)}[I] \;.
\ee 
Here $\tilde{a}^{(T)}$ is given by the r.h.s. of Eq.~(\ref{gfunc_zero_t}) with 
$W_{0}\to W(T)$ and $b^{(T)}$ is chosen such that  $G_{\rm 
ph}(0)=G_{\rm univ}(T/T_K)$. 
The function 
$W(T)$ (with $W(0)=W_0$) 
accounts for the temperature-dependent broadening 
of the step-like features of the zero-temperature xc potentials in Eqs.~(\ref{xcpots_zero_t}). Using 
Eq.~(\ref{idftGph}) together with our ansatz for the xc bias we obtain the 
following condition on the function $b^{(T)}$:
\be
b^{(T)}[N]
= 1 + \frac{c^{(T)}[N]}{\frac{\de \tilde{V}_{\rm xc}}{\de I}
\big|_{\substack{N=1\\I=0}}} 
\left( \frac{1}{G_{\rm univ}} - \frac{1}{G_{{\rm ph},0}} \right),
\label{c_func}
\ee
where $c^{(T)}[1]=1$ and $G_{{\rm ph},0}\equiv G_{{\rm ph},s}(0)$ is the 
KS zero-bias conductance at the ph symmetric point. Since 
$v^{(0)}_{\rm Hxc}[1]=U/2$, from 
Eqs.~(\ref{ks_spectral}) and (\ref{G+-}) we find
$G_{{\rm ph},0}=-\frac{\g^{2}}{2}\int f'(\w)/(\w^{2}+\g^{2}/4)$; thus the term 
in paranthesis is a well defined function of temperature (independent of the functional form of 
$a^{(T)}$). This construction allows for reproducing with 
high accuracy the numerical values of $G_{\rm ph}(0)$ of all the reference 
calculations we compared with. 

\begin{figure}[tbp]
\includegraphics[width=0.47\textwidth]{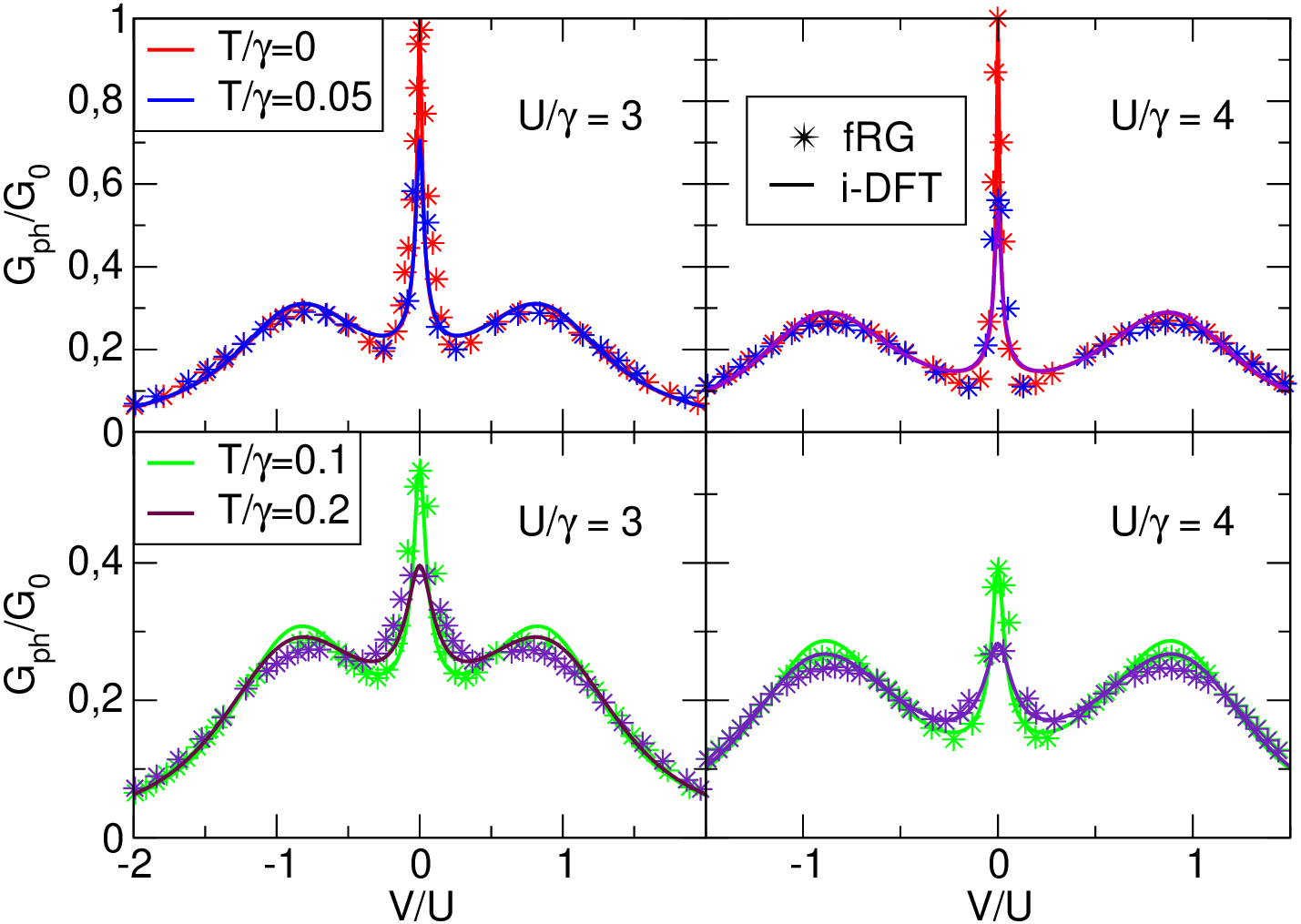}
\caption{Comparison between fRG and i-DFT differential conductances at the 
ph symmetric point 
(in units of the quantum of conductance $G_0=1/\pi$)
as function of bias $V$ for $U/\gamma=3$ (left) and 
$U/\gamma=4$ (right). fRG results from 
Ref.~\protect\citenum{JakobsPletyukhovSchoeller:10}.}
\label{comp_diff_cond_jakobs}
\end{figure}

Although we have not yet specified $c^{(T)}[N]$ for all densities, 
the property $c^{(T)}[1]=1$ is enough to calculate $G_{\rm ph}(V)$ at 
finite bias. Aiming to reproduce  the results 
presented in Ref.~\citenum{JakobsPletyukhovSchoeller:10}, 
we found good agreement 
if we choose the temperature-dependent broadening
\be
W(T) = W_0 \left( 1 + 9 \left(\frac{T}{\g}\right)^2 \right) \;.
\ee
The dependence on the ratio $T/\g$ reflects the physical expectation 
that broadening is dominated by $\g$ at small temperatures and by $T$ 
at high temperatures.
In Fig.~\ref{comp_diff_cond_jakobs} we show the differential conductances at 
the ph symmetric point for $U/\g=3$ and $U/\g=4$ in a large bias window. 
In both cases the i-DFT potentials accurately reproduce the fRG results.

\begin{figure}[tb]
\includegraphics[width=0.47\textwidth]{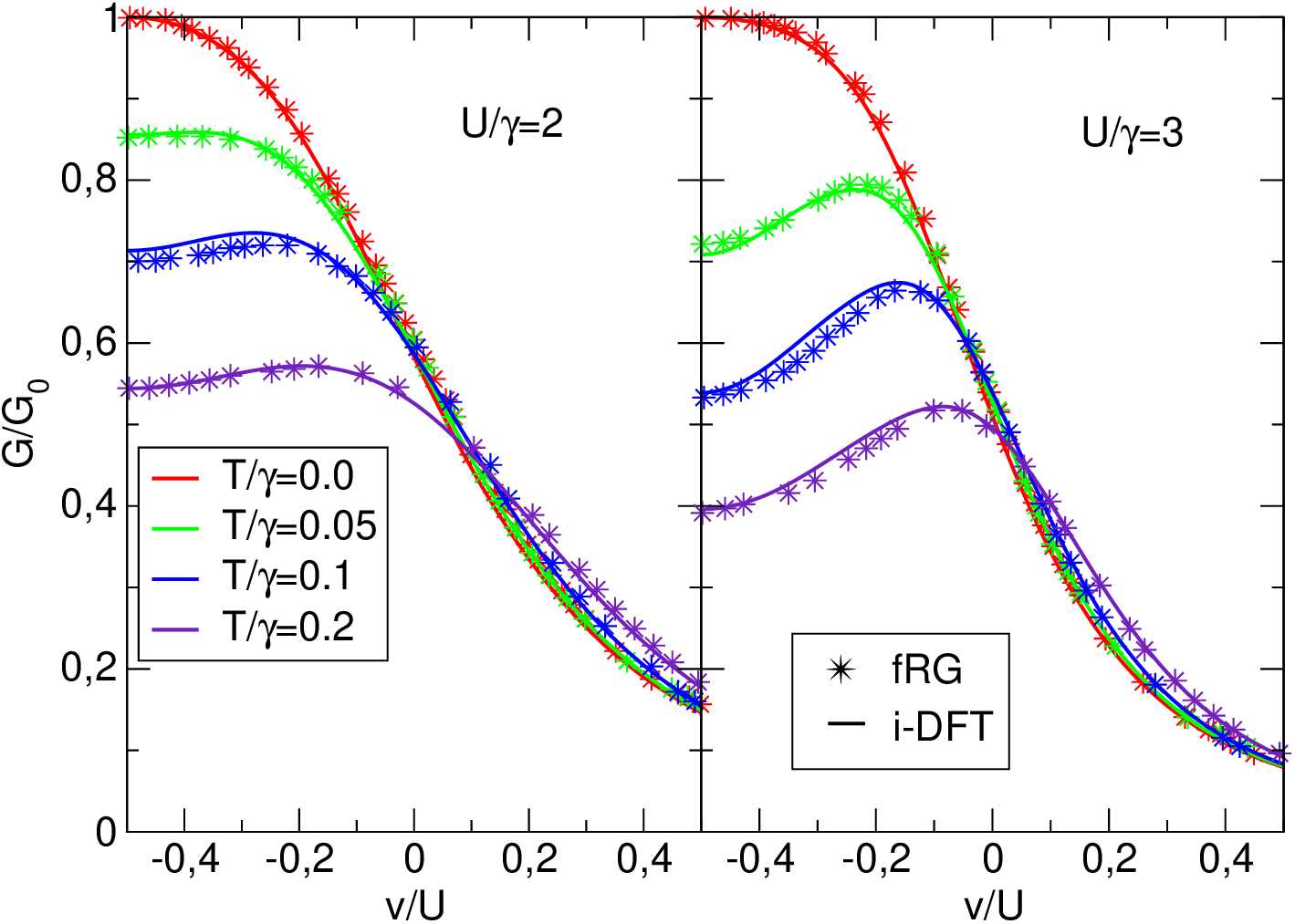}
\caption{Comparison between fRG and i-DFT  zero-bias conductances
(in units of $G_0=1/\pi$)
as function of gate $v$ for $U/\gamma=2$ (left) and 
$U/\gamma=3$ (right). fRG results from
Ref.~\protect\citenum{JakobsPletyukhovSchoeller:10}.}
\label{comp_conduct_jakobs}
\end{figure}

We still need an expression for $c^{(T)}$ which, by ph 
symmetry, is an even function of $(N-1)$. 
Again good agreement between the i-DFT and fRG finite-temperature 
zero-bias conductances is found by choosing
\be
c^{(T)}[N] = 1 + \frac{2}{\pi} \delta(T) 
\arctan\left( \left( \frac{N-1}{\l_{2} W(T)}
\right)^2 \right).
\ee
As the temperature increases 
the Kondo plateau in $G(v)$ is suppressed, and this 
suppression is strongest at the ph symmetric point. The height of the 
resulting ``side peaks'' is controlled  by
\be
\delta(T) = \frac{2}{\pi} \arctan\left( \frac{(U_{c}-U)/\g}{\l_{2} W(T)} 
\right) \;,
\ee
where the values $\l_{2}=3$ and $U_{c}=6 \g$ best fit the fRG results of 
Ref.~\citenum{JakobsPletyukhovSchoeller:10}.
The quality of the i-DFT results for moderate values of $U/\g$
can be appreciated in Fig.~\ref{comp_conduct_jakobs}. 

\begin{figure}[tb]
\includegraphics[width=0.47\textwidth]{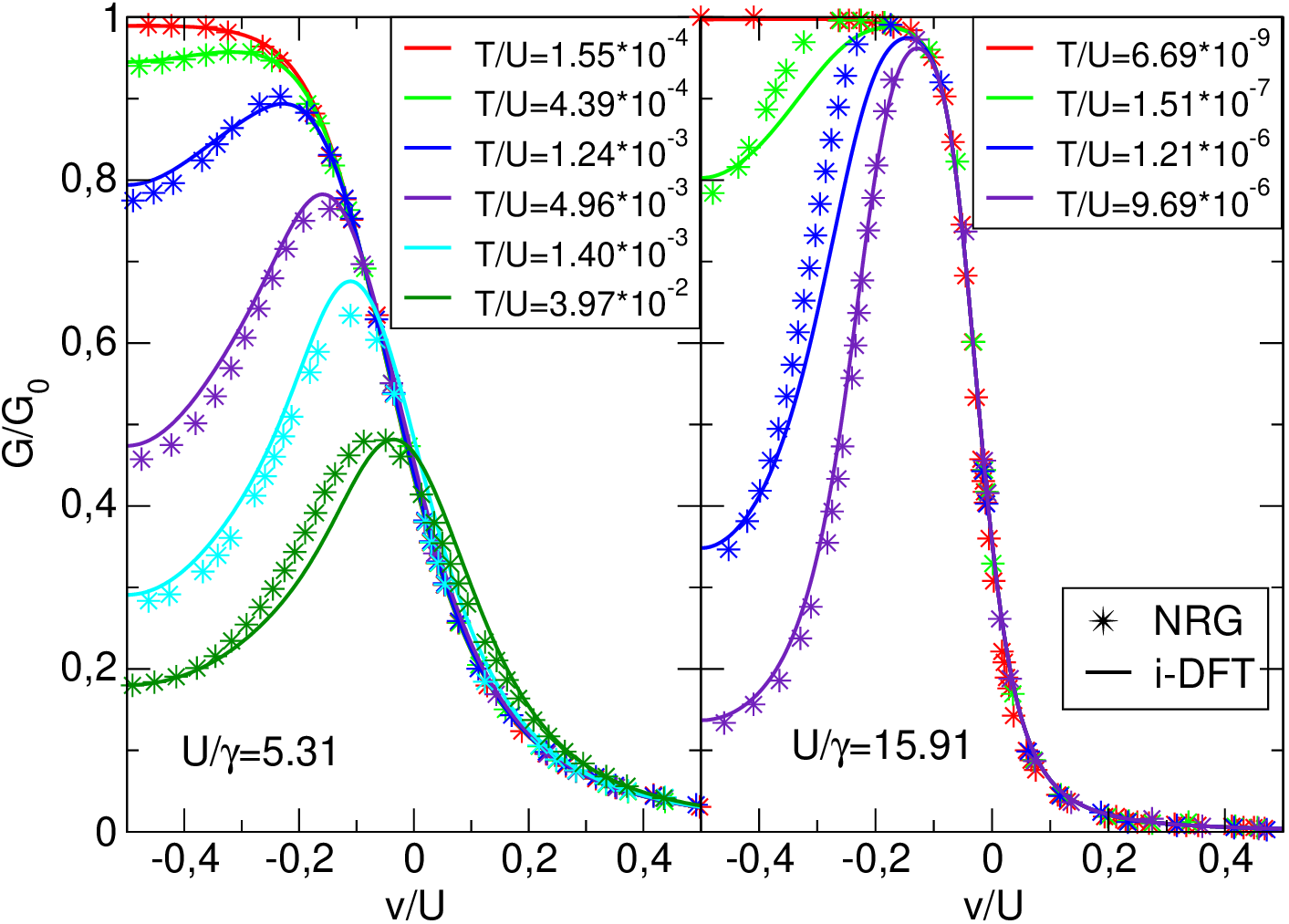}
\caption{Comparison between NRG and i-DFT  zero-bias conductances 
(in units of $G_0=1/\pi$)
as function of gate $v$ for $U/\gamma=5.31$ (left) and 
$U/\gamma=15.91$ (right). NRG results from
Ref.~\protect\citenum{IzumidaSakaiSuzuki:01}.}
\label{comp_conduct_izumida}
\end{figure}

Having fixed the parameters $\l_{1}$, $\l_{2}$ and $U_{c}$ the 
xc potentials can be used to calculate the differential conductance 
for any $U/\gamma$ in a wide range of temperature, gate and 
bias. As a severe test we have analyzed the performance of i-DFT 
in the very strongly correlated regime.
In Fig.~\ref{comp_conduct_izumida} we compare the zero-bias 
conductance of i-DFT and NRG~\cite{IzumidaSakaiSuzuki:01,IzumidaSakai:05}
for several temperatures. Once more the agreement is 
rather satisfactory, only for 
$U/\g=15.91$ and low temperatures the shape of the ``side peaks'' 
is slightly different. 

In conclusion, we demonstrated 
that i-DFT can be used to study the SIAM out of equilibrium, thus disproving 
the common notion that DFT is not suited 
for transport through strongly correlated systems. 
Of course, as the 
construction of the widely used local density approximation in DFT relies 
heavily on accurate xc energies of the uniform electron gas (obtained with, 
e.g., QMC techniques), so the construction of our (H)xc potentials relies 
heavily on accurate conductances of the SIAM obtained with other methods.
However, with an explicit form of the Hxc gate and xc bias the computational 
problem simplifies enormously since the i-DFT equations describe an effectively 
non-interacting system. For any temperature, gate, and interaction 
strength the actual calculation of an $I$-$V$ curve requires 
only negligible computational effort. 
With the (H)xc potentials proposed in this work i-DFT 
becomes a useful and inexpensive method to test and 
benchmark future theoretical techniques in the SIAM. Furthermore, the ideas 
behind the construction of the (H)xc potentials are easily transferable to 
more complicated systems like, e.g., the Constant Interaction Model 
\cite{StefanucciKurth:13,ks.2013,StefanucciKurth:15}, or to 
time-dependent transport (through the adiabatic approximation) 
\cite{kskvg.2010,Verdozzi:08,KarlssonPriviteiraVerdozzi:11,HofmannKuemmel:12,sds.2013,PertsovaStamenovaSanvito:13,FuksMaitra:14}    
and may inform the construction of functionals applicable 
to {\em ab-initio} calculations of correlated materials.

S.K. acknowledges funding by a grant of the "Ministerio de Economia y 
Competividad (MINECO)" (FIS2013-43130-P) and by the 
``Grupos Consolidados UPV/EHU del Gobierno Vasco'' (IT578-13). 
G.S. acknowledges funding by MIUR FIRB Grant No. RBFR12SW0J and EC funding 
through the RISE Co-ExAN (GA644076).

\end{document}